\documentclass[a4paper,11pt]{article}
\usepackage[utf8]{inputenc}
\usepackage{pos}
\newcommand\Tr{\mathrm{Tr}}

\DeclareUnicodeCharacter{2212}{-}

\title{Quantum correction to a new Wilson line-based action for Gluodynamics}
\author[a]{Hiren Kakkad}
\author*[a]{Piotr Kotko}
\author[b]{Anna Stasto}

\affiliation[a]{Faculty of Physics and Applied Computer Science,  AGH University of Krakow,\\
  al. A. Mickiewicza 30, 30-059 Krakow, Poland}

\affiliation[b]{Physics Department, The Pennsylvania State University, \\
104 Davey Lab, University Park, PA 16802, USA}

\emailAdd{piotr.kotko@fis.agh.edu.pl}
\emailAdd{kakkad@agh.edu.pl}

\abstract{We discuss a new classical action that enables efficient computation of the gluonic tree amplitudes but does not contain any triple point vertices. This new formulation is obtained via a canonical transformation of the light-cone Yang-Mills action, with the field transformations based on Wilson line functionals. In addition to MHV vertices, the action contains also N$^k$MHV vertices, where $1\leq k \leq n−4$, and $n$ is the number of external legs. We computed tree-level amplitudes up to 8 gluons and found agreement with standard results. The classical action is however not sufficient to obtain rational parts of amplitudes, in particular the finite amplitudes with all same helicity gluons. In order to systematically develop quantum corrections to this new action, we derive the one-loop effective action, in such a way there are no quantum contributions missing at one loop. }

\FullConference{%
  *** FIXME: Name of conference, ***\\
  *** FIXME: day-day Month YEAR ***\\
  *** FIXME: Location, City, Country ***
}

\begin{document}
\maketitle

%------------------------------------------------------
\section{Introduction}

While most of the recent research on scattering amplitudes focuses on on-shell methods, abandoning the space-time picture altogether in favor of the entirely geometric description \cite{Arkani-Hamed_book_2016}, the present contribution focuses on a new method of computing scattering amplitudes in Yang-Mills theory in a more traditional way -- in terms of off-shell degrees of freedom following from a Lagrangian. Instead of the Yang-Mills fields $A_a^{\mu}$, however, it proves essential to perform a field transformation, such that the new fields will take care of the bulk of diagrams. Such a program was first realized by Mansfield \cite{Mansfield2006}, who applied a canonical field transformation to the transverse components of the gauge fields $A^{\bullet}=(A^{1}+iA^{2})/\sqrt{2}$, $A^{\star}=(A^{1}-iA^{2})/\sqrt{2}$ (the two remaining light-cone components were eliminated by the gauge choice $A^+=0$ and equations of motion) to obtain an action implementing the MHV rules of \cite{Cachazo2004}.
This action, the so-called "MHV action" reads
\begin{equation}
S_{\mathrm{MHV}}\left[{B}^{\bullet}, {B}^{\star}\right]=\int dx^{+}\left(
-\int d^{3}\mathbf{x}\,\mathrm{Tr}\,\hat{B}^{\bullet}\square\hat{B}^{\star} 
+\mathcal{L}_{--+}+\dots+\mathcal{L}_{--+\dots+}+\dots\right)\,,\label{eq:MHV_action}
\end{equation}
where $x^{+}$ is the light-cone time and $\mathbf{x}\equiv\left(x^{-},x^{\bullet},x^{\star}\right)$. We use the 'double-null' coordinates defined as 
$v^{+}=v\cdot\eta$, $v^{-}=v\cdot\tilde{\eta}$, 
$v^{\bullet}=v\cdot\varepsilon_{\bot}^{+}$,  
$v^{\star}=v\cdot\varepsilon_{\bot}^{-}$
with $\eta=\left(1,0,0,-1\right)/\sqrt{2}$, $\tilde{\eta}=\left(1,0,0,1\right)/\sqrt{2}$,
and 
$\varepsilon_{\perp}^{\pm}=\frac{1}{\sqrt{2}}\left(0,1,\pm i,0\right)$. 
%In these coordinates $\square=2(\partial_+\partial_- - \partial_{\bullet}\partial_{\star})$. 
For the fields, we use $\hat{B}=B_at^a$ where $t^a$ are the color generators satisfying  $\left[t^{a},t^{b}\right]=i\sqrt{2}f^{abc}t^{c}$. Owing to this normalization, our coupling constant is re-scaled as $g\rightarrow g/\sqrt{2}$. Above, $\mathcal{L}_{--+\dots+}$ represents the n-point MHV vertex with the following explicit form in the momentum space
\begin{multline}
\mathcal{L}_{--+\dots+}=\int d^{3}\mathbf{p}_{1}\dots d^{3}\mathbf{p}_{n}\delta^{3}\left(\mathbf{p}_{1}+\dots+\mathbf{p}_{n}\right)\,
\widetilde{\mathcal{V}}_{--+\dots+}^{b_{1}\dots b_{n}}\left(\mathbf{p}_{1},\dots,\mathbf{p}_{n}\right)
\\ \widetilde{B}_{b_{1}}^{\star}\left(x^+;\mathbf{p}_{1}\right)\widetilde{B}_{b_{2}}^{\star}\left(x^+;\mathbf{p}_{2}\right)\widetilde{B}_{b_{3}}^{\bullet}\left(x^+;\mathbf{p}_{3}\right)\dots\widetilde{B}_{b_{n}}^{\bullet}\left(x^+;\mathbf{p}_{n}\right)
\,,
\label{eq:MHV_n_point}
\end{multline}
where $\mathbf{p}_i\equiv\left(p_i^{+},p_i^{\bullet},p_i^{\star}\right)$ and 
\begin{equation}
\widetilde{\mathcal{V}}_{--+\dots+}^{b_{1}\dots b_{n}}\left(\mathbf{p}_{1},\dots,\mathbf{p}_{n}\right)= \frac{(-g)^{n-2}}{(n-2)!}  \left(\frac{p_{1}^{+}}{p_{2}^{+}}\right)^{2}
\frac{\widetilde{v}_{21}^{\star 4}}{\widetilde{v}_{1n}^{\star}\widetilde{v}_{n\left(n-1\right)}^{\star}\widetilde{v}_{\left(n-1\right)\left(n-2\right)}^{\star}\dots\widetilde{v}_{21}^{\star}} \mathrm{Tr}\left(t^{b_1}\dots t^{b_n}\right)
\,.
\label{eq:MHV_vertex}
\end{equation}
The symbols ${\widetilde v}^{\star}_{ij}$ and ${\widetilde v}_{ij}$ 
are defined as 
\begin{equation}
   \widetilde{v}^{\star}_{ij}=
    p_i^+\left(\frac{p_{j}^{\bullet}}{p_{j}^{+}}-\frac{p_{i}^{\bullet}}{p_{i}^{+}}\right) , \qquad \widetilde{v}_{ij}=
    p_i^+\left(\frac{p_{j}^{\star}}{p_{j}^{+}}-\frac{p_{i}^{\star}}{p_{i}^{+}}\right)\, .
\label{eq:vtilde}
\end{equation}
When $p_i$, $p_j$ are on-shell ${\widetilde v}^{\star}_{ij}$ and  ${\widetilde v}_{ij}$  are 
proportional to commonly used spinor products $\left<ij\right>$, $[ij]$.

%------------------------------------------------------
\section{The new Wilson line-based action}

\begin{figure}
    \centering
    \includegraphics[width=8cm]{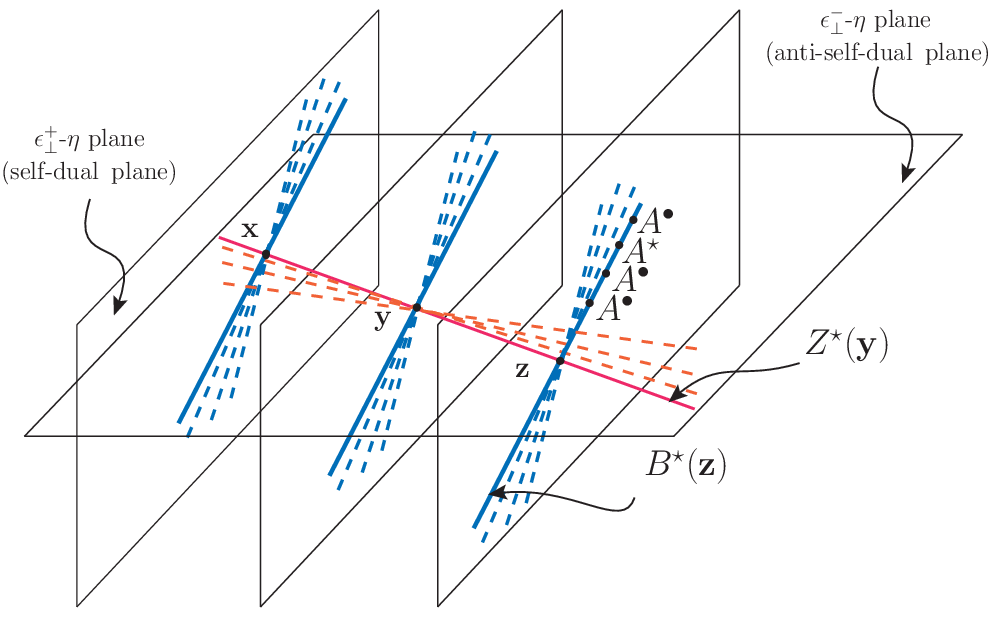}
    \caption{\small
    The red line represents $Z^{\star}$ as a straight infinite Wilson line of $B^{\star}$ lying on the $\varepsilon_{\perp}^{-\, \mu }- \eta^{\mu} $ plane. The dashed line represents the variation of the slope $\alpha$ in Eq.~\eqref{eq:Zstar_WL}. The blue line represents $B^{\star}$ as a straight infinite Wilson line of $A^{\bullet}$ and  $A^{\star}$ fields on the $\varepsilon_{\perp}^{-\, \mu }- \eta^{\mu} $ plane \cite{Kakkad2020}. This figure was taken from \cite{Kakkad:2021uhv}.}
    \label{fig:geometry}
\end{figure}

In \cite{Kakkad:2021uhv}, we derived a new action that goes beyond the MHV action Eq.~\eqref{eq:MHV_action} and provides even more efficient computation of the pure gluonic amplitudes. The new action is derived via a canonical transformation of the light-cone Yang-Mills action, but unlike the transformation leading to MHV action, this transformation maps the kinetic term and both the triple gluon vertices $(++-)$, $(+--)$ of the Yang-Mills action to solely the kinetic term in the new action. However, as demonstrated in \cite{Kakkad:2021uhv}, the same new action could also be derived by canonically transforming the MHV action Eq.~\eqref{eq:MHV_action} via the following 
 \begin{equation}
\int d^{3}\mathbf{x}\, \mathrm{Tr}\,\hat{B}^{\bullet}\left(x^+;\mathbf{x}\right)\square\hat{B}^{\star}\left(x^+;\mathbf{x}\right)\,+\,\mathcal{L}_{--+}[B^{\bullet},B^{\star}]
\,\, \longrightarrow \,\,
\int d^{3}\mathbf{y}\, \mathrm{Tr}\,\hat{Z}^{\bullet}\left(x^+;\mathbf{y}\right)\square\hat{Z}^{\star}\left(x^+;\mathbf{y}\right)
\label{eq:BtoZtransform}
\end{equation}
It maps the kinetic term and the $(+--)$ triple gluon vertex in the MHV action to the kinetic term in the new action. The solution of the above transformation read
\begin{equation}
    Z^{\star}_a[B^{\star}](x)=\int_{-\infty}^{\infty}d\alpha\,\mathrm{Tr}\left\{ \frac{1}{2\pi g}t^{a}\partial_{-}\, \mathbb{P}\exp\left[ig\int_{-\infty}^{\infty}\! ds\, \varepsilon_{\alpha}^{-}\cdot \hat{B}\left(x+s\varepsilon_{\alpha}^{-}\right)\right]\right\}\,,
    \label{eq:Zstar_WL}
    \end{equation}
where $\varepsilon_{\alpha}^{-\, \mu} = \varepsilon_{\perp}^{-\, \mu }- \alpha \eta^{\mu} $,  and 
\begin{equation}
    Z_a^{\bullet}[B^{\bullet},B^{\star}](x) = 
    \int\! d^3\mathbf{y} \,
     \left[ \frac{\partial^2_-(y)}{\partial^2_-(x)} \,
     \frac{\delta Z^{\star}_a[B^{\star}](x^+;\mathbf{x})}{\delta {B}_c^{\star}(x^+;\mathbf{y})} \right] 
     {B}_c^{\bullet}(x^+;\mathbf{y})
      \, .
      \label{eq:Zbul_WL}
\end{equation}
Eq.~\eqref{eq:Zstar_WL} expresses $Z^{\star}$ field as a straight infinite Wilson line of the negative helicity field $B^{\star}$ in MHV action lying on the $\varepsilon_{\perp}^{-\, \mu }- \eta^{\mu} $ plane along $\varepsilon_{\alpha}^{-\, \mu}$ where $\alpha$ is the slope parameter which gets integrated over. Eq.~\eqref{eq:Zbul_WL} expresses $Z^{\bullet}$ field via a similar Wilson line where one of the $B^{\star}$ field has been replaced by a $B^{\bullet}$ field somewhere on the line. Interestingly, the fields in the MHV action $\{B^{\bullet}, B^{\star} \}$ themselves have a Wilson line type structure in terms of the fields $\{A^{\bullet}, A^{\star} \}$ in the Yang-Mills action lying on the perpendicular plane $\varepsilon_{\perp}^{+\, \mu }- \eta^{\mu}$ \cite{Kotko2017, Kakkad2020}. Thus, in terms of gauge fields, $\{Z^{\bullet}, Z^{\star} \}$ fields have a geometric structure of intersecting Wilson lines as shown in Figure \ref{fig:geometry}.

Inverting Eq.~\eqref{eq:Zstar_WL}-\eqref{eq:Zbul_WL}, in momentum space we get
\begin{equation}
    \widetilde{B}^{\star}_a(x^+;\mathbf{P}) = \sum_{n=1}^{\infty} 
    \int d^3\mathbf{p}_1\dots d^3\mathbf{p}_n \, \overline{\widetilde{\Psi}}\,^{a\{b_1\dots b_n\}}_n(\mathbf{P};\{\mathbf{p}_1,\dots ,\mathbf{p}_n\}) \prod_{i=1}^n\widetilde{Z}^{\star}_{b_i}(x^+;\mathbf{p}_i)\,,
    \label{eq:BstarZ_exp}
\end{equation}
\begin{equation}
    \widetilde{B}^{\bullet}_a(x^+;\mathbf{P}) = \sum_{n=1}^{\infty} 
    \int d^3\mathbf{p}_1\dots d^3\mathbf{p}_n \, \overline{\widetilde \Omega}\,^{a b_1 \left \{b_2 \cdots b_n \right \}}_{n}(\mathbf{P}; \mathbf{p_1} ,\left \{ \mathbf{p_2} , \dots ,\mathbf{p_n} \right \}) \widetilde{Z}^{\bullet}_{b_1}(x^+;\mathbf{p}_1)\prod_{i=2}^n\widetilde{Z}^{\star}_{b_i}(x^+;\mathbf{p}_i)\,,
    \label{eq:BbulletZ_exp}
\end{equation}
\begin{equation}
    \overline{\widetilde \Psi}\,^{a \left \{b_1 \cdots b_n \right \}}_{n}(\mathbf{P}; \left \{\mathbf{p}_{1},  \dots ,\mathbf{p}_{n} \right \}) =- (-g)^{n-1} \,\,  
    \frac{{\widetilde v}_{(1 \cdots n)1}}{{\widetilde v}_{1(1 \cdots n)}} \, 
    \frac{\delta^{3} (\mathbf{p}_{1} + \cdots +\mathbf{p}_{n} - \mathbf{P})\,\,  \mathrm{Tr} (t^{a} t^{b_{1}} \cdots t^{b_{n}})}{{\widetilde v}_{21}{\widetilde v}_{32} \cdots {\widetilde v}_{n(n-1)}}  
      \, ,
    \label{eq:psiBar_kernel}
\end{equation}
\begin{equation}
    \overline{\widetilde \Omega}\,^{a b_1 \left \{b_2 \cdots b_n \right \}}_{n}(\mathbf{P}; \mathbf{p}_{1} , \left \{ \mathbf{p}_{2} , \dots ,\mathbf{p}_{n} \right \} ) = n \left(\frac{p_1^+}{p_{1\cdots n}^+}\right)^2 \overline{\widetilde \Psi}\,^{a b_1 \cdots b_n }_{n}(\mathbf{P}; \mathbf{p}_{1},  \dots ,\mathbf{p}_{n}) \, .
    \label{eq:omegaBar_kernel}
\end{equation}
Substituting Eqs.~\eqref{eq:BstarZ_exp}-\eqref{eq:BbulletZ_exp} to the MHV action Eq.~\eqref{eq:MHV_action} we get the new action (which we dub as the "Z-field action" here onwards)
\begin{align}
S\left[Z^{\bullet},Z^{\star}\right] =\int dx^{+} \Bigg\{  
\mathcal{L}_{-+} 
 & + \mathcal{L}_{--++}+ \mathcal{L}_{--+++}+\mathcal{L}_{--++++} + \dots \nonumber \\
& + \mathcal{L}_{---++}+ \mathcal{L}_{---+++}+\mathcal{L}_{---++++} + \dots \nonumber \\
& \,\, \vdots \nonumber \\
& + \mathcal{L}_{---\dots -++}+ \mathcal{L}_{---\dots -+++}+\mathcal{L}_{---\dots -++++}+ \dots
\Bigg\}
\,,\label{eq:Z_action}
\end{align}
where
\begin{equation}
    \mathcal{L}_{\underbrace{-\,\cdots\,-}_{m}\underbrace{+ \,\cdots\, +}_{n-m}}= 
   \int\!d^{3}\mathbf{p}_{1}\dots d^{3}\mathbf{p}_{n} \,\, \mathcal{U}^{b_1 \dots b_{n}}_{-\dots-+\dots+}\left(\mathbf{p}_{1},\cdots \mathbf{p}_{n}\right) 
   \prod_{i=1}^{m}Z^{\star}_{b_i} (x^+;\mathbf{p}_{i})
   \prod_{j=1}^{n-m}Z^{\bullet}_{b_j} (x^+;\mathbf{p}_{j}) \, ,
   \label{eq:Z_vertex_lagr_mom}
\end{equation}
represents a generic $n$-point interaction vertex with $m$ minus helicity legs. Above we have introduced
\begin{equation}
    \mathcal{U}_{-\dots-+\dots+}^{b_{1}\dots b_{n}}\left(\mathbf{p}_{1},\dots,\mathbf{p}_{n}\right)= \!\!\sum_{\underset{\text{\scriptsize permutations}}{\text{noncyclic}}}
 \mathrm{Tr}\left(t^{b_1}\dots t^{b_n}\right)
 \mathcal{U}\left(1^-,\dots,m^-,(m+1)^+,\dots,n^+\right)
\,,
\label{eq:Zvertex_color_decomp}
\end{equation}
with
\begin{multline}
    \mathcal{U}\left(1^-,\dots,m^-,(m\!+\!1)^+,\dots,n^+\right) = 
    \sum_{p=0}^{m-2}\sum_{q=p+1}^{m-1}\sum_{r=q+1}^{m}\\
    \mathcal{V}_{\mathrm{MHV}}\left(\,[p\!+\!1,\dots,q]^-,[q\!+\!1,\dots,r]^-,[r\!+\!1,\dots,m\!+\!1]^+,(m\!+\!2)^+,\dots,(n\!-\!1)^+,[n,1,\dots,p]^+\right) \\
    \overline{ \Omega}\left(n^+,1^-,\dots,p^-\right) 
    \overline{ \Psi}\left((p\!+\!1)^-,\dots,q^-\right)  
    \overline{ \Psi}\left((q\!+\!1)^-,\dots,r^-\right)  
    \overline{ \Omega}\left((r\!+\!1)^-,\dots,m^-,(m\!+\!1)^+\right) \,,
    \label{eq:Zver_MHV}
\end{multline}
where $\mathcal{V}_{\mathrm{MHV}}\left(1^-,2^-,3^+\dots, n^+\right)$, $
    \overline{ \Psi}\left(1^-,\dots,n^-\right) $, and $\overline{ \Omega}\left(1^+,2^-,\dots,n^-\right) $ are the the color-ordered MHV vertex Eq.~\eqref{eq:MHV_vertex}, and the kernels in Eqs.~\eqref{eq:psiBar_kernel}-\eqref{eq:omegaBar_kernel} respectively. Above, $[i,\dots,j] \equiv \mathbf{p}_{i} + \dots +\mathbf{p}_{j}$. We validated the above expression by computing amplitudes up to 8 points and found agreement with standard results \cite{Dixon:2010ik}. The maximum number of diagrams for 8-point NNMHV amplitude was 13,  which is way less than what one gets in the MHV action.
%------------------------------------------------------
\section{Quantum Corrections}

The Z-field action is incomplete when it comes to computing loop amplitudes; the amplitudes where all the gluons have either the same helicities i.e. $(+ \dots +)$, $(- \dots -)$ or one of them has a different helicity i.e. $(- + \dots +)$, $(+ - \dots -)$ are zero in our action Eq.~\eqref{eq:Z_action}. These amplitudes indeed vanish at the tree level but are non-zero at one loop \cite{Bern:1991aq,Kunszt_1994}. 
This fact has been attributed to quantum anomalies in the self-dual (anti-self dual) sector of the Yang-Mills theory \cite{Bardeen1996,Chattopadhyay:2020oxe,Monteiro:2022nqt}. 

In order to control the quantum corrections at one loop level, we construct the one-loop effective action in such a way that the self-dual and anti-self dual sectors are incorporated explicitly. This can be done in two ways. The most straightforward way is to derive the one-loop effective action for the Yang-Mills theory first and then apply the field redefinitions. The drawback of that approach is that the vertices of the Z-field theory are not explicit in the loop. Second approach is to transform the Yang-Mills action, together with the linear, source dependent term, which after the transformation is performed encodes the missing contributions. In what follows, we shall focus on the latter method.
We begin with the generating functional for the full Green's function for the Yang-Mills action and apply the canonical transformation that derives the Z-field action
\begin{equation}
    \mathcal{Z}[J]=\int[dA]\, e^{i\left(S_{\mathrm{YM}}[A] + \int\!d^4x\, \Tr \hat{J}_j(x) \hat{A}^j(x)\right) } \longrightarrow \int[dZ]\, e^{i\left(S[Z] + \int\!d^4x\, \Tr \hat{J}_j(x) \hat{A}^j[Z](x)\right) } \,.
   \label{eq:SZ_genr}
\end{equation}
Above, $\hat{J}_j$ represents the external current coupled with the action, $j = \bullet, \star$,  $S[Z]$ is the Z-field action, and $\hat{A}^j[Z]$ is the solution of the canonical transformation that derives it from the Yang-Mills action \cite{Kakkad:2021uhv}. Notice, we do not introduce new currents when turning the Yang-Mills action into the Z-field action. This is crucial to keep the contributions from the self-dual and anti-self dual sector of the Yang-Mills. Now, we expand the terms in the exponent around the classical field configuration ${\hat Z}^i_{c}=\left\{{\hat Z}_{c}^{\bullet}(x), {\hat Z}_{c}^{\star}(x) \right\}$ up to second order
\begin{multline}
    S[Z] + \int\!d^4x\, \Tr \hat{J}_i(x) \hat{A}^i[Z](x)  
    = S[Z_c] + \int\!d^4x\, \Tr \hat{J}_i(x)\hat{A}^i[Z_c](x) \\ + \int\!d^4x\,\Tr\left(\hat{Z}^i(x)-\hat{Z}_c^i(x)\right)
    \left(\frac{\delta S[Z_c]}{\delta \hat{Z}^i(x)}+\int\!d^4y\,{\hat J}_k(y)\frac{\delta \hat{A}^k[Z_c](y)}{\delta \hat{Z}^i(x)}\right) 
    +\frac{1}{2}\int\!d^4xd^4y\,\Tr\left(\hat{Z}^i(x)-\hat{Z}_c^i(x)\right)\\
    \left(\frac{\delta^2 S[Z_c]}{\delta\hat{Z}^i(x)\delta\hat{Z}^j(y)}+\int\!d^4z\,{\hat J}_k(z)\frac{\delta^2 \hat{A}^k[Z_c](z)}{\delta \hat{Z}^i(x)\delta\hat{Z}^j(y)}\right)\left(\hat{Z}^j(y)-\hat{Z}_c^j(y)\right) \,.
    \label{eq:Der_olea_sz}
\end{multline}
Higher order terms are necessary for quantum corrections beyond one-loop, thus we drop them. Performing the Gaussian functional integral we get
\begin{equation}
   \mathcal{Z}[J] \approx  
   \exp\left\{iS\left[Z_c\left[J\right]\right] 
    + i\int\!d^4x\, \Tr \hat{J}_i(x)\hat{A}^i[Z_c[J]](x) -\frac{1}{2}\Tr\ln\mathrm{M}[J] \right\} \,,
    \label{eq:det_SZln}
\end{equation}
where
$\mathrm{M}[J]=\mathrm{M}^{\text{Z-field}}[J]+\mathrm{M}^{\text{src}}[J]$, 
with
 \begin{equation}
\mathrm{M}^{\text{Z-field}}_{IK}[J]  =  \left(\begin{matrix}
     \frac{\delta^2 S[Z_c]}{\delta Z^{\bullet I}\delta Z^{\star K}} 
     &
     \frac{\delta^2 S[Z_c]}{\delta Z^{\bullet I}\delta Z^{\bullet K}}\\ \\
\frac{\delta^2 S[Z_c]}{\delta Z^{\star I}\delta Z^{\star K}}
      &
      \frac{\delta^2 S[Z_c]}{\delta Z^{\star I}\delta Z^{\bullet K}} 
\end{matrix}\right) \,,
\label{eq:M_Zfield}
\end{equation}
and
\begin{equation}
\mathrm{M}^{\text{src}}_{IK}[J] = \left(\begin{matrix}
    J_{\star L}\frac{\delta^2 A^{\star L}[Z_c]}{\delta Z^{\bullet I}\delta Z^{\star K}} + J_{\bullet L}\frac{\delta^2 A^{\bullet L}[Z_c]}{\delta Z^{\bullet I}\delta Z^{\star K}} 
     & J_{\star L}\frac{\delta^2 A^{\star L}[Z_c]}{\delta Z^{\bullet I}\delta Z^{\bullet K}} +  J_{\bullet L}\frac{\delta^2 A^{\bullet L}[Z_c]}{\delta Z^{\bullet I}\delta Z^{\bullet K}}\\ \\
 J_{\star L}\frac{\delta^2 A^{\star L}[B_c]}{\delta Z^{\star I}\delta Z^{\star K}} + J_{\bullet L}\frac{\delta^2 A^{\bullet L}[B_c]}{\delta Z^{\star I}\delta Z^{\star K}}
      &J_{\star L}\frac{\delta^2 A^{\star L}[B_c]}{\delta Z^{\star I}\delta Z^{\bullet K}} + J_{\bullet L}\frac{\delta^2 A^{\bullet L}[B_c]}{\delta Z^{\star I}\delta Z^{\bullet K}}
\end{matrix}\right) \,.
\label{eq:M_Zsrc}
\end{equation}
Above, we introduced the collective indices $I,J,K \dots$ which run over the color, position, etc. Eq.~\eqref{eq:det_SZln} represents the generating functional for the Green's function up to one loop. The loop diagrams originate from the trace of the log term. The One-Loop Effective Action $\Gamma[Z_c]$ is obtained via the Legendre transform of the generating functional for the connected Green's function 
\begin{equation}
   \Gamma[Z_c] = W[J] - \int\!d^4x\, \Tr \hat{J}_i(x)\hat{A}^i[Z_c[J]](x) \,, \quad \mathrm{where} \quad W[J] = -i \ln \left[ \mathcal{Z}[J]\right]\,.
\end{equation}
Substituting Eq.~\eqref{eq:det_SZln} above, we get
\begin{equation}
   \Gamma[Z_c] = S\left[Z_c\right] 
    + i\frac{1}{2}\Tr\ln\mathrm{M}[J[Z_c]]\,.
    \label{eq:olea_zth}
\end{equation}
Notice, the sources in the matrix $\mathrm{M}[J]$ have been substituted as functional of the fields $\hat{Z}^i_c$. This is done by rewriting the classical equations of motion as follows (see \cite{kakkad2023scattering} for the derivation)
\begin{equation}
  \left.\frac{\delta S_{\mathrm{YM}}[A]}{\delta A^{\star L}}\right|_{A=A[Z_c]}= -J_{\star L}\,, \quad\quad
    \left. \frac{\delta S_{\mathrm{YM}}[A]}{\delta A^{\bullet L}}\right|_{A=A[Z_c]}= - J_{\bullet L} \,,
    \label{eq:J_currZ}
\end{equation}
where $S_{\mathrm{YM}}$ is light-cone Yang-Mills action. The log term in Eq.~\eqref{eq:olea_zth} can be evaluated using
\begin{equation}
    \Tr\ln\mathrm{M}= \Tr\ln\mathrm{M}_{\bullet\star} +
    \Tr\ln\left(\mathrm{M}_{\star\bullet}
    - \mathrm{M}_{\star\star}\mathrm{M}^{-1}_{\bullet\star}\mathrm{M}_{\bullet\bullet}\right) \,.
    \label{eq:lnM_decomp}
\end{equation}
This relation follows from the determinant of block matrices with non-commuting blocks. Above, $\mathrm{M}_{\bullet\star}=\mathrm{M}_{\star\bullet}$ are the diagonal blocks,  $\mathrm{M}_{\star\star}$, $\mathrm{M}_{\bullet\bullet}$ are respectively the bottom-left and top-right blocks of the matrix $\mathrm{M}[J[Z_c]]$.

Interestingly, the one-loop effective action Eq.~\eqref{eq:olea_zth} with the complicated log term can be simplified to (modulo a field independent volume divergent factor) the following
\begin{multline}
   \Gamma[Z_c] = S[Z_c] 
    + \frac{i}{2} \Tr\ln \left[ \frac{\delta^2 S_{\mathrm{YM}}[A]}
    {\delta A^{\star I}\delta A^{\bullet K}} \, \frac{\delta^2 S_{\mathrm{YM}}[A]}
    {\delta A^{\star K}\delta A^{\bullet J}} \right. \\
    \left.- \frac{\delta^2 S_{\mathrm{YM}}[A]}
    {\delta A^{\star I}\delta A^{\bullet K}} \, \frac{\delta^2 S_{\mathrm{YM}}[A}
    {\delta A^{\star K}\delta A^{\star L}} \left( \frac{\delta^2 S_{\mathrm{YM}}[A]}
    {\delta A^{\bullet L}\delta A^{\star M}} \right)^{-1} \frac{\delta^2 S_{\mathrm{YM}}[A]}
    {\delta A^{\bullet M}\delta A^{\bullet J}}\right]_{A=A[Z_c]}\,.
    \label{eq:OLEA_Z}
\end{multline}
The above action can be directly obtained by starting with the one-loop effective action for the Yang-Mills and then performing the transformations that derive the Z-field action (in fact, we used this approach to successfully develop one-loop corrections to the MHV action Eq.~\eqref{eq:MHV_action} \cite{Kakkad_2022}). The  difference between Eq.~\eqref{eq:OLEA_Z} and Eq.~\eqref{eq:olea_zth} is in the log term. The former uses Yang-Mills vertices in the loop and the solutions $\hat{A}^j[Z]$ account for all the tree-level connections originating from both $(+ + -)$ and $(--+)$ triple gluon vertices. On the other hand, in Eq.~\eqref{eq:OLEA_Z}, the Z-field vertices are explicit in the loop. Which method is ultimately more efficient is yet to be determined. Eq.~\eqref{eq:olea_zth} has been used explicitly to compute all 4-point one-loop amplitudes: $(+ + ++)$, $(----)$, $(+ + +-)$, $(---+)$, and $(--++)$ (see \cite{kakkad2023scattering}). This also implies that there are no missing loop contributions in Eq.~\eqref{eq:OLEA_Z} making both effective actions one-loop complete. An additional advantage of both actions is that they require much less diagrams for computing higher multiplicity one-loop amplitude when compared with the Yang-Mills action (see \cite{kakkad2023scattering} for further details).

%------------------------------------------------------
\section{Acknowledgements}
H.K. is supported by the National Science Centre, Poland grant no. 2021/41/N/ST2/02956. P.K. is supported by the National Science Centre, Poland grant no. 2018/31/D/ST2/02731.  A.M.S. is supported  by the U.S. Department of Energy Grant 
 DE-SC-0002145 and  in part by  National Science Centre in Poland, grant 2019/33/B/ST2/02588.

 \small 
 
\bibliographystyle{JHEP}
\bibliography{library}

\end{document}